\PassOptionsToPackage{unicode}{hyperref}
\PassOptionsToPackage{hyphens}{url}
\PassOptionsToPackage{dvipsnames,svgnames,x11names}{xcolor}
\documentclass[
]{article}
\usepackage{amsmath,amssymb}
\usepackage{lmodern}
\usepackage{iftex}
\ifPDFTeX
  \usepackage[T1]{fontenc}
  \usepackage[utf8]{inputenc}
  \usepackage{textcomp} 
\else 
  \usepackage{unicode-math}
  \defaultfontfeatures{Scale=MatchLowercase}
  \defaultfontfeatures[\rmfamily]{Ligatures=TeX,Scale=1}
\fi
\IfFileExists{upquote.sty}{\usepackage{upquote}}{}
\IfFileExists{microtype.sty}{
  \usepackage[]{microtype}
  \UseMicrotypeSet[protrusion]{basicmath} 
}{}
\makeatletter
\@ifundefined{KOMAClassName}{
  \IfFileExists{parskip.sty}{%
    \usepackage{parskip}
  }{
    \setlength{\parindent}{0pt}
    \setlength{\parskip}{6pt plus 2pt minus 1pt}}
}{
  \KOMAoptions{parskip=half}}
\makeatother
\usepackage{xcolor}
\providecommand{\tightlist}{%
  \setlength{\itemsep}{0pt}\setlength{\parskip}{0pt}}
\NewDocumentCommand\citeproctext{}{}
\NewDocumentCommand\citeproc{mm}{%
  \begingroup\def\citeproctext{#2}\cite{#1}\endgroup}
\makeatletter
 \let\@cite@ofmt\@firstofone
 \def\@biblabel#1{}
 \def\@cite#1#2{{#1\if@tempswa , #2\fi}}
\makeatother
\newlength{\cslhangindent}
\newlength{\csllabelwidth}
\newenvironment{CSLReferences}[2] 
 {\begin{list}{}{%
  \setlength{\itemindent}{0pt}
  \setlength{\leftmargin}{0pt}
  \setlength{\parsep}{0pt}
  \ifodd #1
   \setlength{\leftmargin}{\cslhangindent}
   \setlength{\itemindent}{-1\cslhangindent}
  \fi
  \setlength{\itemsep}{#2\baselineskip}}}
 {\end{list}}
\usepackage{calc}

\ifLuaTeX
\usepackage[bidi=basic]{babel}
\else
\usepackage[bidi=default]{babel}
\fi
\babelprovide[main,import]{american}

\def\languageshorthands#1{}
\ifLuaTeX
  \usepackage{selnolig}  
\fi
\IfFileExists{bookmark.sty}{\usepackage{bookmark}}{\usepackage{hyperref}}
\IfFileExists{xurl.sty}{\usepackage{xurl}}{} 
\hypersetup{
  pdftitle={unxt: A Python package for unit-aware computing with JAX},
  pdfauthor={Nathaniel Starkman, Adrian M. Price-Whelan, Jake Nibauer},
  pdflang={en-US},
  colorlinks=true,
  linkcolor={Maroon},
  filecolor={Maroon},
  citecolor={Blue},
  urlcolor={Blue},
  pdfcreator={LaTeX via pandoc}}

\title{unxt: A Python package for unit-aware computing with JAX}

\definecolor{c53baa1}{RGB}{83,186,161}
\definecolor{c202826}{RGB}{32,40,38}
\def \rorglobalscale {0.1}
\newcommand{\rorlogo}{%
\begin{tikzpicture}[y=1cm, x=1cm, yscale=\rorglobalscale,xscale=\rorglobalscale, every node/.append style={scale=\rorglobalscale}, inner sep=0pt, outer sep=0pt]
  \begin{scope}[even odd rule,line join=round,miter limit=2.0,shift={(-0.025, 0.0216)}]
    \path[fill=c53baa1,nonzero rule,line join=round,miter limit=2.0] (1.8164, 3.012) -- (1.4954, 2.5204) -- (1.1742, 3.012) -- (1.8164, 3.012) -- cycle;
    \path[fill=c53baa1,nonzero rule,line join=round,miter limit=2.0] (3.1594, 3.012) -- (2.8385, 2.5204) -- (2.5172, 3.012) -- (3.1594, 3.012) -- cycle;
    \path[fill=c53baa1,nonzero rule,line join=round,miter limit=2.0] (1.1742, 0.0669) -- (1.4954, 0.5588) -- (1.8164, 0.0669) -- (1.1742, 0.0669) -- cycle;
    \path[fill=c53baa1,nonzero rule,line join=round,miter limit=2.0] (2.5172, 0.0669) -- (2.8385, 0.5588) -- (3.1594, 0.0669) -- (2.5172, 0.0669) -- cycle;
    \path[fill=c202826,nonzero rule,line join=round,miter limit=2.0] (3.8505, 1.4364).. controls (3.9643, 1.4576) and (4.0508, 1.5081) .. (4.1098, 1.5878).. controls (4.169, 1.6674) and (4.1984, 1.7642) .. (4.1984, 1.8777).. controls (4.1984, 1.9719) and (4.182, 2.0503) .. (4.1495, 2.1132).. controls (4.1169, 2.1762) and (4.0727, 2.2262) .. (4.0174, 2.2635).. controls (3.9621, 2.3006) and (3.8976, 2.3273) .. (3.824, 2.3432).. controls (3.7505, 2.359) and (3.6727, 2.367) .. (3.5909, 2.367) -- (2.9676, 2.367) -- (2.9676, 1.8688).. controls (2.9625, 1.8833) and (2.9572, 1.8976) .. (2.9514, 1.9119).. controls (2.9083, 2.0164) and (2.848, 2.1056) .. (2.7705, 2.1791).. controls (2.6929, 2.2527) and (2.6014, 2.3093) .. (2.495, 2.3487).. controls (2.3889, 2.3881) and (2.2728, 2.408) .. (2.1468, 2.408).. controls (2.0209, 2.408) and (1.905, 2.3881) .. (1.7986, 2.3487).. controls (1.6925, 2.3093) and (1.6007, 2.2527) .. (1.5232, 2.1791).. controls (1.4539, 2.1132) and (1.3983, 2.0346) .. (1.3565, 1.9436).. controls (1.3504, 2.009) and (1.3351, 2.0656) .. (1.3105, 2.1132).. controls (1.2779, 2.1762) and (1.2338, 2.2262) .. (1.1785, 2.2635).. controls (1.1232, 2.3006) and (1.0586, 2.3273) .. (0.985, 2.3432).. controls (0.9115, 2.359) and (0.8337, 2.367) .. (0.7519, 2.367) -- (0.1289, 2.367) -- (0.1289, 0.7562) -- (0.4837, 0.7562) -- (0.4837, 1.4002) -- (0.6588, 1.4002) -- (0.9956, 0.7562) -- (1.4211, 0.7562) -- (1.0118, 1.4364).. controls (1.1255, 1.4576) and (1.2121, 1.5081) .. (1.2711, 1.5878).. controls (1.2737, 1.5915) and (1.2761, 1.5954) .. (1.2787, 1.5991).. controls (1.2782, 1.5867) and (1.2779, 1.5743) .. (1.2779, 1.5616).. controls (1.2779, 1.4327) and (1.2996, 1.3158) .. (1.3428, 1.2113).. controls (1.3859, 1.1068) and (1.4462, 1.0176) .. (1.5237, 0.944).. controls (1.601, 0.8705) and (1.6928, 0.8139) .. (1.7992, 0.7744).. controls (1.9053, 0.735) and (2.0214, 0.7152) .. (2.1474, 0.7152).. controls (2.2733, 0.7152) and (2.3892, 0.735) .. (2.4956, 0.7744).. controls (2.6016, 0.8139) and (2.6935, 0.8705) .. (2.771, 0.944).. controls (2.8482, 1.0176) and (2.9086, 1.1068) .. (2.952, 1.2113).. controls (2.9578, 1.2253) and (2.9631, 1.2398) .. (2.9681, 1.2544) -- (2.9681, 0.7562) -- (3.3229, 0.7562) -- (3.3229, 1.4002) -- (3.4981, 1.4002) -- (3.8349, 0.7562) -- (4.2603, 0.7562) -- (3.8505, 1.4364) -- cycle(0.9628, 1.7777).. controls (0.9438, 1.7534) and (0.92, 1.7357) .. (0.8911, 1.7243).. controls (0.8623, 1.7129) and (0.83, 1.706) .. (0.7945, 1.7039).. controls (0.7588, 1.7015) and (0.7252, 1.7005) .. (0.6932, 1.7005) -- (0.4839, 1.7005) -- (0.4839, 2.0667) -- (0.716, 2.0667).. controls (0.7477, 2.0667) and (0.7805, 2.0643) .. (0.8139, 2.0598).. controls (0.8472, 2.0553) and (0.8768, 2.0466) .. (0.9025, 2.0336).. controls (0.9282, 2.0206) and (0.9496, 2.0021) .. (0.9663, 1.9778).. controls (0.9829, 1.9534) and (0.9914, 1.9209) .. (0.9914, 1.8799).. controls (0.9914, 1.8362) and (0.9819, 1.8021) .. (0.9628, 1.7777) -- cycle(2.6125, 1.3533).. controls (2.5889, 1.2904) and (2.5553, 1.2359) .. (2.5112, 1.1896).. controls (2.4672, 1.1433) and (2.4146, 1.1073) .. (2.3529, 1.0814).. controls (2.2916, 1.0554) and (2.2228, 1.0427) .. (2.1471, 1.0427).. controls (2.0712, 1.0427) and (2.0026, 1.0557) .. (1.9412, 1.0814).. controls (1.8799, 1.107) and (1.8272, 1.1433) .. (1.783, 1.1896).. controls (1.7391, 1.2359) and (1.7052, 1.2904) .. (1.6817, 1.3533).. controls (1.6581, 1.4163) and (1.6465, 1.4856) .. (1.6465, 1.5616).. controls (1.6465, 1.6359) and (1.6581, 1.705) .. (1.6817, 1.7687).. controls (1.7052, 1.8325) and (1.7388, 1.8873) .. (1.783, 1.9336).. controls (1.8269, 1.9799) and (1.8796, 2.0159) .. (1.9412, 2.0418).. controls (2.0026, 2.0675) and (2.0712, 2.0804) .. (2.1471, 2.0804).. controls (2.223, 2.0804) and (2.2916, 2.0675) .. (2.3529, 2.0418).. controls (2.4143, 2.0161) and (2.467, 1.9799) .. (2.5112, 1.9336).. controls (2.5551, 1.8873) and (2.5889, 1.8322) .. (2.6125, 1.7687).. controls (2.636, 1.705) and (2.6477, 1.6359) .. (2.6477, 1.5616).. controls (2.6477, 1.4856) and (2.636, 1.4163) .. (2.6125, 1.3533) -- cycle(3.8015, 1.7777).. controls (3.7825, 1.7534) and (3.7587, 1.7357) .. (3.7298, 1.7243).. controls (3.701, 1.7129) and (3.6687, 1.706) .. (3.6333, 1.7039).. controls (3.5975, 1.7015) and (3.5639, 1.7005) .. (3.5319, 1.7005) -- (3.3226, 1.7005) -- (3.3226, 2.0667) -- (3.5547, 2.0667).. controls (3.5864, 2.0667) and (3.6192, 2.0643) .. (3.6526, 2.0598).. controls (3.6859, 2.0553) and (3.7155, 2.0466) .. (3.7412, 2.0336).. controls (3.7669, 2.0206) and (3.7883, 2.0021) .. (3.805, 1.9778).. controls (3.8216, 1.9534) and (3.8301, 1.9209) .. (3.8301, 1.8799).. controls (3.8301, 1.8362) and (3.8206, 1.8021) .. (3.8015, 1.7777) -- cycle;
  \end{scope}
\end{tikzpicture}
}

\usepackage[affil-it]{authblk}
\usepackage{orcidlink}
\author[1%
  \ensuremath\mathparagraph]{Nathaniel Starkman%
    \,\orcidlink{0000-0003-3954-3291}\,%
    }
\author[2%
  ]{Adrian M. Price-Whelan%
    \,\orcidlink{0000-0003-0872-7098}\,%
    }
\author[3%
  ]{Jake Nibauer%
    \,\orcidlink{0000-0001-8042-5794}\,%
    }

\affil[1]{Brinson Prize Fellow at Kavli Institute for Astrophysics and
Space Research, Massachusetts Institute of Technology, USA%
    \,\protect\href{https://ror.org/042nb2s44}{\protect\rorlogo}\,%
  }
\affil[2]{Center for Computational Astrophysics, Flatiron Institute,
USA%
    \,\protect\href{https://ror.org/00sekdz59}{\protect\rorlogo}\,%
  }
\affil[3]{Department of Physics, Princeton University, USA%
    \,\protect\href{https://ror.org/00hx57361}{\protect\rorlogo}\,%
  }
\affil[$\mathparagraph$]{Corresponding author: %
}
\date{15 November 2024}

\begin{document}
\maketitle

\section{Summary}\label{summary}

\texttt{unxt} is a Python package for unit-aware computing with JAX
(\citeproc{ref-jax:18}{Bradbury et al., 2018}), which is a
high-performance numerical computing library that enables automatic
differentiation and just-in-time compilation to accelerate code
execution on multiple compute architectures. \texttt{unxt} is built on
top of \texttt{quax} (\citeproc{ref-quax:23}{Kidger, 2023}), which
provides a framework for building array-like objects that can be used
with JAX. \texttt{unxt} extends \texttt{quax} to provide support for
unit-aware computing using the \texttt{astropy.units} package
(\citeproc{ref-astropy:13}{Astropy Collaboration et al., 2013},
\citeproc{ref-astropy:22}{2022}) as a units backend. \texttt{unxt}
provides seamless integration of physical units into high performance
numerical computations, significantly enhancing the capabilities of JAX
for scientific applications.

The primary purpose of \texttt{unxt} is to facilitate unit-aware
computations in JAX, ensuring that operations involving physical
quantities are handled correctly and consistently. This is crucial for
avoiding errors in scientific calculations, such as those that could
lead to significant consequences like the infamous Mars Climate Orbiter
incident (\citeproc{ref-nasa:98}{NASA}). \texttt{unxt} is designed to be
intuitive, easy to use, and performant, allowing for a straightforward
implementation of units into existing JAX codebases.

\texttt{unxt} is accessible to researchers and developers, providing a
user-friendly interface for defining and working with units and unit
systems. It supports both static and dynamic definitions of unit
systems, allowing for flexibility in various computational environments.
Additionally, \texttt{unxt} leverages multiple dispatch to enable deep
interoperability with other libraries, currently \texttt{astropy}, and
to support custom array-like objects in JAX. This extensibility makes
\texttt{unxt} a powerful tool for a wide range of scientific and
engineering applications, where unit-aware computations are essential.

\section{Statement of Need}\label{statement-of-need}

JAX is a powerful tool for high-performance numerical computing,
offering features such as automatic differentiation, just-in-time
compilation, and support for sharding computations across multiple
devices. It excels in providing unified interfaces to various compute
architectures, including CPUs, GPUs, and TPUs, to accelerate code
execution (\citeproc{ref-jax:18}{Bradbury et al., 2018}). However, JAX
operates primarily on ``pure'' arrays, which means it lacks support to
define custom array-like objects, including those that can handle units,
and to use those use those objects in within the JAX ecosystem. While
JAX can handle PyTrees with some pre-programmed support and the ability
to register additional support, the operations it performs are still
fundamentally array-based. This limitation poses a challenge for
scientific applications that require handling of physical units.

Astropy has been an invaluable resource for the scientific community,
with over 10,000 citations to its initial paper and more than 2,000
citations to its 2022 paper (\citeproc{ref-astropy:13}{Astropy
Collaboration et al., 2013}, \citeproc{ref-astropy:22}{2022}). One of
the foundational sub-packages within Astropy is \texttt{astropy.units},
which provides robust support for units and quantities, enabling the
propagation of units through NumPy functions. This functionality ensures
that scientific calculations involving physical quantities are handled
correctly and consistently. However, despite JAX's numpy-like API, it
does not support the same level of extensibility, and
\texttt{astropy.units} cannot be directly extended to work with JAX.
This gap highlights the need for a solution that integrates the powerful
unit-handling capabilities of Astropy with the high-performance
computing features of JAX.

\texttt{unxt} addresses this gap by providing a function-oriented
framework---consistent with the style of JAX---for handling units and
dimensions, with an object-oriented front-end that will be familiar to
users of \texttt{astropy.units}. By leveraging \texttt{quax},
\texttt{unxt} defines a \texttt{Quantity} class that seamlessly
integrates with JAX functions. This integration is achieved by providing
a comprehensive set of overrides for JAX primitives, ensuring that users
can utilize the \texttt{Quantity} class without needing to worry about
the underlying JAX interfacing. This design allows users to perform
unit-aware computations effortlessly, maintaining the high performance
and flexibility that JAX offers while ensuring the correctness and
consistency of operations involving physical quantities.

\section{Related Works}\label{related-works}

\texttt{unxt} is designed to be extensible to other unitful-computation
libraries. The \texttt{unxt} package is not intended to replace these
libraries, but rather to provide a JAX-optimized frontend. Some
prominent libraries include:

\begin{itemize}
\tightlist
\item
  \texttt{astropy.units} (\citeproc{ref-astropy:13}{Astropy
  Collaboration et al., 2013}, \citeproc{ref-astropy:22}{2022}). The
  \texttt{unxt} package currently uses the unit conversion framework
  from \texttt{astropy.units} package in its backend, providing a more
  flexible front-end interface and particularly JAX-compatible Quantity
  classes for doing array computations with units.
\item
  \texttt{unyt} (\citeproc{ref-unyt:2018}{Goldbaum et al., 2018}). The
  \texttt{unyt} library is a popular Python package for unit-aware
  computations. It provides \texttt{Quantity} classes that work with (at
  time of writing) \texttt{numpy} (\citeproc{ref-numpy:2020}{Harris et
  al., 2020}) and \texttt{dask} (\citeproc{ref-dask:2016}{Dask
  Development Team, 2016}) arrays.
\item
  \texttt{pint} (\citeproc{ref-pint}{Grecco, 2012}). The \texttt{pint}
  library is a popular Python package for unit-aware computations. It
  provides \texttt{Quantity} classes that work with many array types,
  but not \texttt{jax} (at time of writing).
\end{itemize}

\section{Acknowledgements}\label{acknowledgements}

Support for this work was provided by The Brinson Foundation through a
Brinson Prize Fellowship grant.

The authors thank the Astropy collaboration and many contributors for
their work on \texttt{astropy}, which has been invaluable to the
scientific community. Members of the \texttt{unxt} development team are
also core developers and maintainers of the \texttt{astropy.units}
package, and we had \texttt{astropy.units} as our guiding star while
developing \texttt{unxt}. The authors also thank Dan Foreman-Mackey for
useful discussions, and the attendees of the 2024 JAXtronomy workshop at
the Center for Computational Astrophysics at the Flatiron Institute. We
also extend our gratitude to Patrick Kidger for his valuable
communications and guidance on using \texttt{quax} to ensure seamless
integration of \texttt{unxt} with \texttt{jax}.

\section*{References}\label{references}
\addcontentsline{toc}{section}{References}

\protect\phantomsection\label{refs}

\end{document}